\documentclass[aps,prl,twocolumn,showpacs,superscriptaddress,groupedaddress,floatfix]{revtex4-1}
\usepackage{graphicx}
\usepackage{amsmath}
\usepackage{amssymb}
\usepackage{dcolumn}
\usepackage{natbib}
\usepackage{dsfont}
\usepackage{latexsym}
\usepackage{rotating}
\usepackage{color}
\usepackage{latexsym}
\usepackage{bbm}
\usepackage{subfigure}
\usepackage{float}
\usepackage{epsfig}
\usepackage{psfrag}
\usepackage{bm}
\usepackage{amsthm}
\usepackage{eucal}
\usepackage{mathrsfs}
\usepackage{braket}
\usepackage{psfrag}
\usepackage{enumerate}
\usepackage{longtable}
\usepackage{subfigure}
\usepackage{bm}
\usepackage{hyperref}

\hypersetup{
colorlinks=true,final=true,
        linkcolor=red,
        citecolor=blue,
        filecolor=blue,
        urlcolor=blue,
}

\usepackage{dcolumn}
\usepackage{bm}
\usepackage{dcolumn}
\usepackage{bm}
\newcommand{\be}{\begin{equation}}
\newcommand{\ee}{\end{equation}}
\newcommand{\bn}{\begin{eqnarray}}
\newcommand{\en}{\end{eqnarray}}
\begin{document}

\title{Hidden Fermi Liquidity and Topological Criticality in the Finite Temperature Kitaev Model}

\author{Subhasree Pradhan$^{1}$}\email{spradhan@phy.iitkgp.ernet.in}
\author{M. S. Laad$^{2}$}\email{mslaad@imsc.res.in} 
\author{Avijeet Ray$^{3}$}
\author{T. Maitra$^{3}$}
\author{A. Taraphder$^{1,4}$}

\affiliation{$^{1}$Department of Physics, Indian Institute of Technology, Kharagpur, Kharagpur 721302, India}
\affiliation{$^{2}$Institute of Mathematical Sciences, Taramani, Chennai 600113, India, and Homi Bhabha National Institute, Mumbai, India
}
\affiliation{$^{3}$Department of Physics, Indian Institute of Technology Roorkee, Uttarakhand, India}
\affiliation{$^{4}$Centre for Theoretical Studies, Indian Institute of Technology Kharagpur, Kharagpur 721302, India}

\begin{abstract}
{ The fate of exotic spin liquid states with fractionalized excitations at finite temperature ($T$) is of great interest, since signatures of fractionalization manifest in finite-temperature ($T$) dynamics in real systems, above the tiny magnetic ordering scales. Here, we study a Jordan-Wigner fermionized Kitaev spin liquid at finite $T$ employing combined Exact diagonalization and Monte Carlo simulation methods. We uncover $(i)$ checkerboard or stripy-ordered flux crystals depending on density of flux, and $(ii)$ establish, surprisingly, that: $(a)$ the finite-$T$ version of the $T=0$ transition from a gapless to gapped phases in the Kitaev model is a Mott transition of the fermions, belonging to the two-dimensional Ising universality class. These transitions correspond to a topological transition between a string condensate and a dilute closed string state $(b)$ the Mott ``insulator'' phase is a precise realization of Laughlin's gossamer (here, $p$-wave) superconductor (g-SC), and $(c)$ the Kitaev Toric Code phase (TC) is a {\it fully} Gutzwiller-projected fermi sea of JW fermions. These findings establish the finite-$T$ QSL phases in the $d = 2$ to be {\it hidden} Fermi liquid(s) of neutral fermions.   
}
\end{abstract}

\maketitle

The birth of the exactly solvable Kitaev model (KM)~\cite{kitaev} exhibiting quantum spin liquid (QSL) behaviour with fractionalized excitations has led to a spurt of activity in various contexts~\cite{khaliullin,jilla,jja}. Whilst studied numerically~\cite{trebst1}, a physically satisfying picture of finite-temperature responses in terms of changes in the underlying spectrum of elementary excitations of the QSL state remains elusive. In a QSL, the fractionalized excitations are expected to manifest as a broad continuum of spin excitations in various scattering experiments~\cite{moessner} (inelastic neutron, two-magnon Raman, resonant inelastic X-ray probes). A complication is that small residual (e.g, Heisenberg) couplings frustrate such a quest at very low $T$ in real systems, and thus extension of ground state investigations to $T$ larger than the small ordering scales, where the novel spin excitations should reveal themselves, is crucial toward establishing spin liquidity in practice.

  The KM is attractive because of its exact solvability in $d=2,3$. The fermionized Kitaev model can be transformed to a $p$-wave SC, and hence is exactly solvable at $T=0$ (because of the flux-free condition), but this does not hold at finite $T$. Specifically, finite but not-too low $T$ excites finite density of ``fluxes'': in a Jordan-Wigner (JW) fermionization, the KM maps on to the spinless Falicov-Kimball model (FKM) with a finite $p$-wave BCS term for the JW fermions~\cite{chen} (see supplementary). The $Z_{2}$ flux variables are, remarkably, recast as immobile spinless fermions. This has no exact solution in $d = 2,3$, so finite $T$ investigations have received much less attention in literature. Yoshitake {\it et al}~\cite{motome} have recently used cluster-dynamical mean-field theory (CDMFT) to investigate the effect of fluctuating fractionalized spins on finite-$T$ magnetic fluctuation responses. The FK-like interaction between the mobile and immobile Majorana fermions on the $zz$-links of the honeycomb lattice lead to peculiar responses even at low $T$ where the {\it static} spin correlations saturate. At low $T\simeq 0.01J$, the specific heat shows a distinct peak that sharpens up near the gapless-to-gapful transition in the KM: since any conventional magnetic order is ruled out, thanks to the exact solution at $T=0$, this novel feature could arise from self-organization of the thermally excited $Z_{2}$-fluxes into an Ising-like checkerboard pattern corresponding to the famous staggered ($\pi$) flux phase~\cite{affleck} without any conventional order in RVB-like spin liquid scenario.

 This is an interesting but hitherto uninvestigated possibility: depending on their density, the static but annealed $Z_{2}$ fluxes in the FKM can also ``self-organize'' themselves into ordered, checkerboard and/or stripe-like ordered phases, or remain in a flux-disordered state at finite $T$: this is indeed well-known to occur in the spinless-FK model in $d = 2$~\cite{Kennedy}. On the other hand, the $p$-wave BCS and hopping terms involving mobile JW fermions will compete with the local Hubbard-like coupling. As the parameter $J_{z}/J$ (with $J=J_{x}=J_{y}$) is varied, one also generically expects a ``Mott transition'' involving gap opening in the fermionic spectrum to occur. In this context, the issues we focus on here are the following: $(1)$ given that JW-fermionic bilinear in the KM correspond to complicated non-local {\it strings} of original spin operators~\cite{nussinov}, the ``Mott'' transition in the fermions corresponds to string condensate(s) in the original KM. Does the JW ``insulator'' then correspond to a ``normal'' phase of small closed strings?~\cite{wen} $(2)$ What is the universality class of such a topological string condensate-to-dilute closed string transition? $(3)$ since a finite and sizable $p$-wave SC term exists in the KM, the ``Mott insulator'' of JW fermions should have ``preformed'' $p$-wave pairing at any finite $J_{z}$? If so, one might  envisage a connection to Anderson's resonating valence bond (RVB) idea in the sense of a Mott insulator of JW fermions with preformed $p$-wave pairs. Whether such a connection can be more rigorously characterized in terms of Laughlin's gossamer SC is an enticing idea we investigate here.
  
\vspace{0.2cm}

Consider the Kitaev model:

\be
H=-\sum_{\alpha}J_{\alpha}\sum_{<i,j>}S_{i}^{\alpha}S_{j}^{\alpha}
\ee

\noindent Using the $d = 2$ Jordan-Wigner mapping~\cite{hu}, $H$(see supplementary informations) can be brought to the following instructive form (this also bares the {\it exact} emergent low-dimensional gauge symmetries, as pointed out by Fradkin et al.~\cite{fradkin}):

\begin{align}
H_{K}=&\sum_{q}[\epsilon_{q}c_{q}^{\dag}c_{q}+\frac{i\Delta_{q}}{2}(c_{q}^{\dag}c_{-q}^{\dag}+h.c.)]
\nonumber\\&+\frac{J_{z}}{4}\sum_{i}(2n_{i\alpha}-1)(2c_{i}^{\dag}c_{i}-1)
\end{align}
 with an additional ``spin-triplet'' pairing term along with the FK terms, whence the spin liquid ground state of $H_K$ now appears as a $p$-wave BCS superfluid of JW fermions. Here, we re-express the $Z_{2}$ flux on $zz$ bonds in the KM as $\alpha_{i}'=(2n_{i\alpha}-1)=iB_{i1}B_{i2}=\pm 1$, which is thus an {\it annealed} gauge field variable on the center of each zz-links of the honeycomb Kitaev model (when $n_{i,\alpha}=0,1$ for all $i$). This transforms the original spin model on a honeycomb lattice to a two- ``orbital" spinless model on an effective square lattice.

In JW-fermion language, the now {\it dynamical} $Z_{2}$ flux can, by itself, self-organize into stripe-ordered patterns, in analogy with crystal states found in the FK model~\cite{Kennedy}. In the present case, such crystals occurring with a large range of periodic structures have a novel interpretation as flux crystals involving condensation of topological kink dipoles (a condensation of $\langle iB_{i1}B_{i2}\rangle$ on each zz-bond). Remarkably, we find that in the fermionized $H_{K}$, this topological transition manifests itself as checkerboard or stripe ordering of the $\alpha$-fermions. As $J_{z}/J$ increases, one expects a ``Mott-Hubbard''-like phase transition between a gapless spin liquid ($p$-wave BCS) to a gapped ``Mott insulator'' of the JW fermions as $J_{z}/J$ increases. These features are consistent with the observation~\cite{chen} that topological phase transitions can be viewed as conventional orders in a dual (JW or Majorana fermion) representation. We have carefully investigated $(1)-(3)$ above at finite but low $T\simeq 0.01J$~\cite{motome} using a classical Monte Carlo technique based on exact diagonalization, (MC+ED) method that is ideally suitable for FK-like models~\cite{Taraphder}. 
\vspace{0.50cm}
\vspace{0.1em}
\begin{figure}
\centering
\includegraphics[angle=0,,trim=1.0in 1.6in 0in 1.6in,width=1.1\columnwidth]{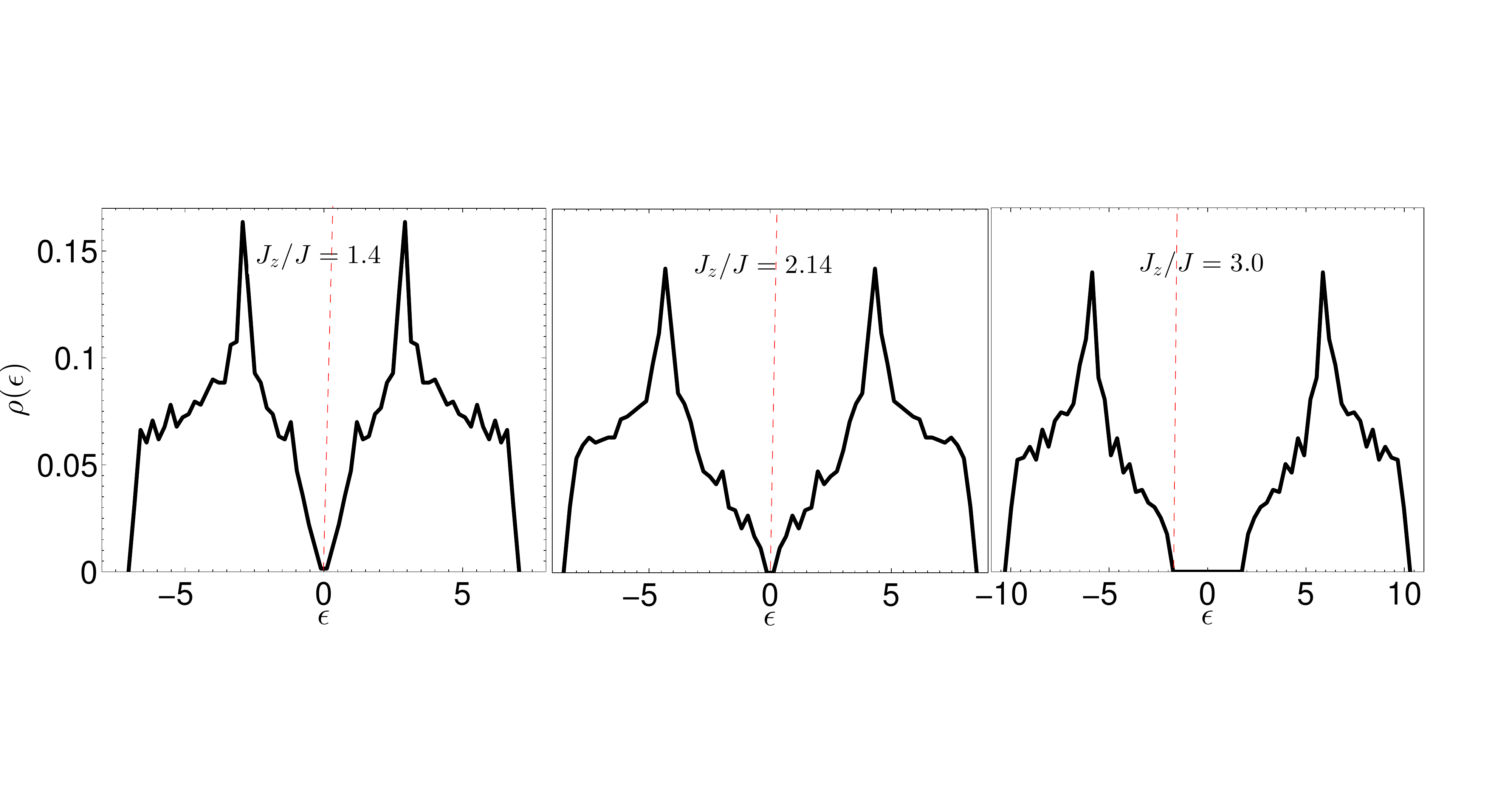}
\caption{(Color online) Density of States plot for $J = 1, Jz = 1.4, 2.14$ and $3.0$.  A phase transition from ``semi-metal''-to-``Mott'' insulator, with {\it both} phases having $p$-wave superfluidity, occurs at $J_{z}^{(c)}/J\simeq 2.2$ (not shown).}
\label{fig1}
\end{figure}

\begin{figure}[ht!]
\centering
\subfigure{\label{f:C11}\epsfig{file=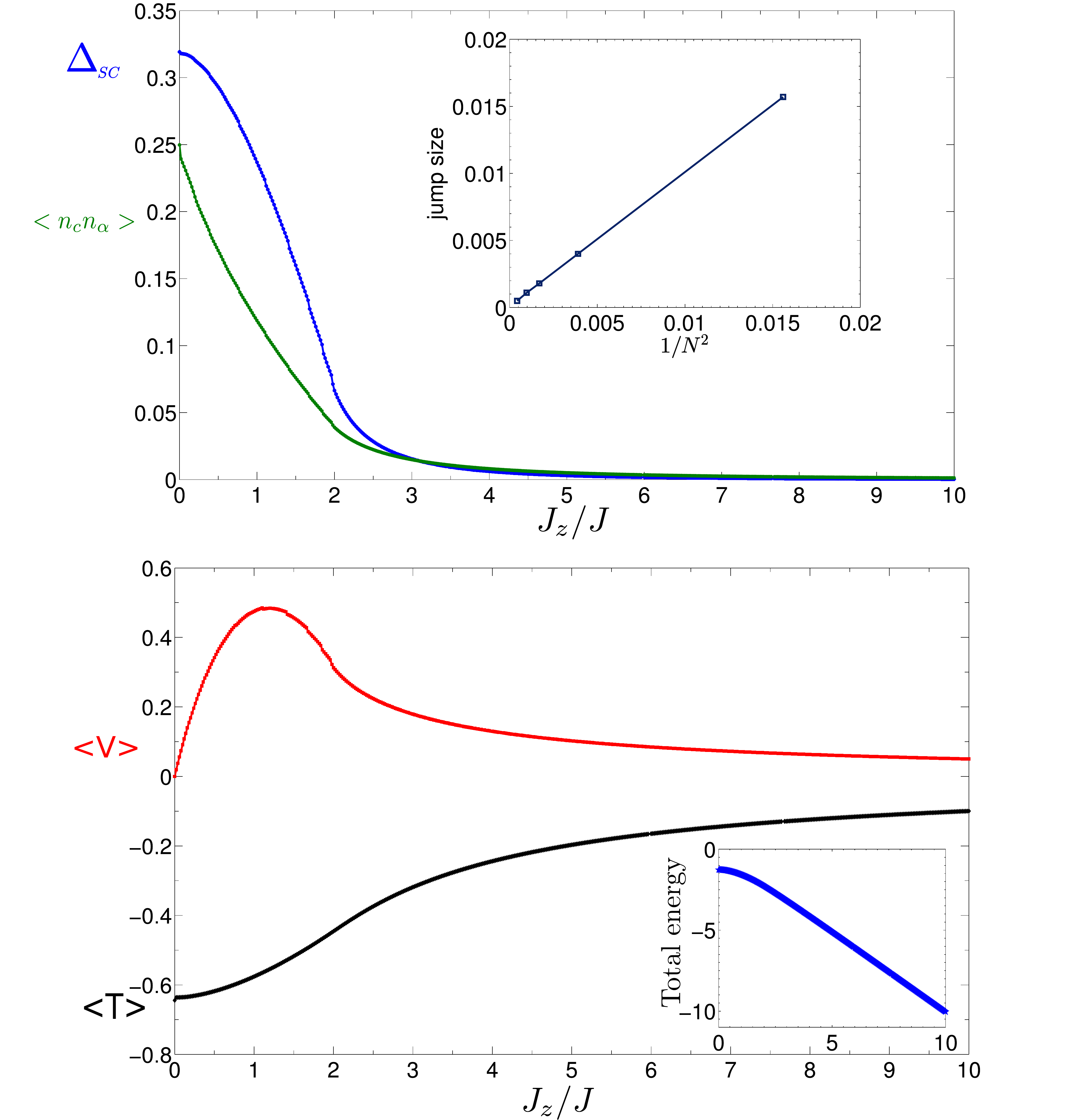,trim=0in 0in 0in 0.0in,clip=true,width=1.0\linewidth}}\hspace{-0.0\linewidth}
\caption{(Color online) Upper panel shows the numerically extracted $p$-wave SC gap amplitude ($\Delta_{sc}=\langle c_{i}^{\dag}c_{j}^{\dag}\rangle$) and the double occupancy, $D=\langle n_{ic}n_{i\alpha}\rangle$, as functions of $J_{z}/J$ (in units of $J$). Lower panel shows average kinetic ($\langle T\rangle$) and potential ($\langle V\rangle$) energies, along with the total energy (inset). The inset in the upper panel shows how the finite-size-induced jumps in $\Delta_{sc},D$ vanish upon lattice size extrapolation to the thermodynamic limit.}
\label{fig2}
\end{figure}

In the upper panel of Fig.~\ref{fig1}, we show results for the local density-of-states (LDOS) of the JW fermions for $\langle n_{ic}\rangle=1/2=\langle n_{i\alpha}\rangle$. Taking $J=1$ as the unit of energy, the $p$-wave BCS state remains stable up to a critical $J_{z}^{(c)}\simeq 2.14$, beyond which a {\it continuous} transition to a gapped ``Mott'' phase of the JW fermions obtains (this is well known to occur at $J_{z}^{(c)}=2.0$ at $T=0$). In the lower panel, we bare the analogy with Hubbard-like physics (in the JW fermion sector) by showing the variation of the double occupancy, $D=\langle n_{ic}n_{i\alpha}\rangle$ and the $p$-wave BCS pair average $\Delta_{ij}=\langle c_{i}^{\dag}c_{j}^{\dag}\rangle$ as a function of $J_{z}$. While a ``Mott'' gap continuously opens in the LDOS at $J_{z}^{(c)}$, it is clear that {\it both}, $D$ and $\Delta_{ij}$ remain sizable even in the gapped phase. Though there are finite lattice-size induced jumps in both, these reduce with increasing lattice size and disappear upon extrapolation to the thermodynamic limit (see inset of Fig.~\ref{fig2}). The first ($D(J_{z})$) is reminiscent of the situation in the Mott transition in the Hubbard model, where $D$ remains finite at the transition, vanishing smoothly as $U\rightarrow\infty$. That $\Delta_{ij}\neq 0$ beyond $J_{z}^{(c)}$ is now interpretable as a {\it preformed} $p$-wave pairing in the JW-fermion Mott insulator: this is a distinguishing feature of the original RVB proposal~\cite{science1987}. We now exploit this insight to propose that a {\it gossamer spin liquid phase} emerges in the Kitaev model at finite $T$.   


\vspace{0.1em}

To facilitate this, we show the average kinetic energy, $\langle T\rangle=\langle c_{i}^{\dag}c_{j}\rangle$ per bond and average potential energy, $\langle V\rangle=\langle n_{ic}n_{i\alpha}\rangle$ for $H_{FK}$ in Fig.~\ref{fig2} for a $36\times36$ lattice. Both exhibit clear changes in slope across the ``Mott'' transition, but no non-analyticities, consistent with a continuous phase transition (in fact, it is the {\it second} derivatives which show a discontinuity). Using these results, it is straightforward to estimate the ``Gutzwiller renormalization factors'' $g_{t}=\langle T\rangle/\langle T\rangle_{0}, g_{ij}=\langle \Delta_{ij}\rangle/\langle \Delta_{ij}\rangle_{0}$ associated with renormalization of kinetic energy and $p$-wave BCS averages as a function of $J_{z}$. Given that both are reduced but still sizable up to $J_{z}\simeq O(5)$, beyond which they smoothly vanish for $J_{z}\geq 8.0$ implies that the ``Mott insulator'' still has sizable double occupancy {\it co-existing} with $p$-wave BCS correlations. Thus, this is precisely a Mott insulator of JW fermions with preformed pairs: in fact, more precisely, since $D (J_{z}>J_{z}^{(c)})$ is not small, we have found that a {\it partially} Gutzwiller projected $p$-wave BCS wavefunction describes this ``Mott'' (gapped QSL in the original KM) state. A direct connection to Laughlin's gossamer (here, $p$-wave) superfluid thus emerges, described by a correlated wavefunction

\be
|\Psi\rangle = \Pi_{\eta}\Pi_{k}(u_{k}+v_{k}c_{k}^{\dag}c_{-k}^{\dag})|JWFS\rangle
\ee
where $|JWFS\rangle$ is the Fermi sea of non-interacting JW fermions, and $\Pi_{\eta}=\Pi_{i}(1-\eta n_{ic}n_{i\alpha})$ is the partial Gutzwiller projector accounting for finite double occupancy in the JW-Mott insulator. In the regime $2.14\leq J_{z}\leq 6.0$, we also see that both $\langle V\rangle$, $g_{ij}$ are severely reduced, attesting to the gossamer state. Given that finite $\langle c_{i}^{\dag}c_{j}^{\dag}\rangle$ is associated with the spin liquid in original spin variables, we christen this state a {\it gossamer spin liquid}. In the Kitaev QSL phase, the Fourier transformed pair average corresponds to a highly non-trivial {\it string} correlator (see Eq.39 of Chen {\it et al.}~\cite{chen} as follows: $P_{k,k'}=\langle c_{k}^{\dag}c_{-k}^{\dag}c_{-k'}c_{k'}\rangle$ with

\be
c_{k}^{\dag}c_{-k}^{\dag}=\sum_{r_{w1},r_{w2}}e^{ik(r_{w2}-r_{w1})}P(r_{w1},r_{w2})
\ee
where ${P(r_{w1},r_{w2}})=[S_{r_{w1}}^{y}(\Pi_{r'<r_{w1}}S_{r'}^{z})-iS_{r_{b1}^{z}}(\Pi_{r'<r_{b1}}S_{r'}^{z})].[S_{r_{w2}}^{y}(\Pi_{r''<r_{w2}}S_{r''}^{z})-iS_{r_{b2}^{z}}(\Pi_{r''<r_{b2}}S_{r''}^{z})]$ is a complicated and highly non-local string of spin operators. Thus, the $p$-wave SC state is a {\it string} condensate, implying that the QSL is a fluid state of proliferated non-local strings. The observation that the $p$-wave condensate is severely reduced by strong but partial Gutzwiller projection in the ``Mott'' insulator for $2.14<J_{z}<6.0$ corresponds to a dominant quantum phase fluctuation dominated ``gossamer'' $p$-wave SC, and indicates a ``thin'' string condensate in spin language. For $J_{z}>6.0$, this condensate is completely destroyed: in JW fermion language, this is a superconductor-insulator transition (SIT), driven by competition between pairing and localization (the source of the latter is strong scattering between JW fermions and the static $Z_{2}$ fluxes which act as intrinsic {\it annealed} ``disorder'' in $H_{FK}$). This insulator without any pairing amplitude is one characterized by {\it total} Gutzwiller projection, since $D$ also vanishes for $J_{z}>8.0$. In the spin language, the gossamer QSL gives way to the toric code (TC) phase of the KM, which is now described by the famous {\it fully} projected (here, $p$-wave) BCS wavefunction of JW fermions. Remarkably, we thus find that at finite $T$ with a macroscopic $Z_{2}$ flux density, the Kitaev QSL first undergoes a transition to a gossamer QSL, followed by a subsequent transition to the TC phase. The first manifests as a continuous ``Mott'' transition with a co-existing finite pairing correlation, while the second manifests as a phase-fluctuation dominated continuous SI transition. This is one of the central messages of our paper.

\vspace{0.1em}
\begin{figure}[ht!]
\centering
\subfigure{\label{f:C11}\epsfig{file=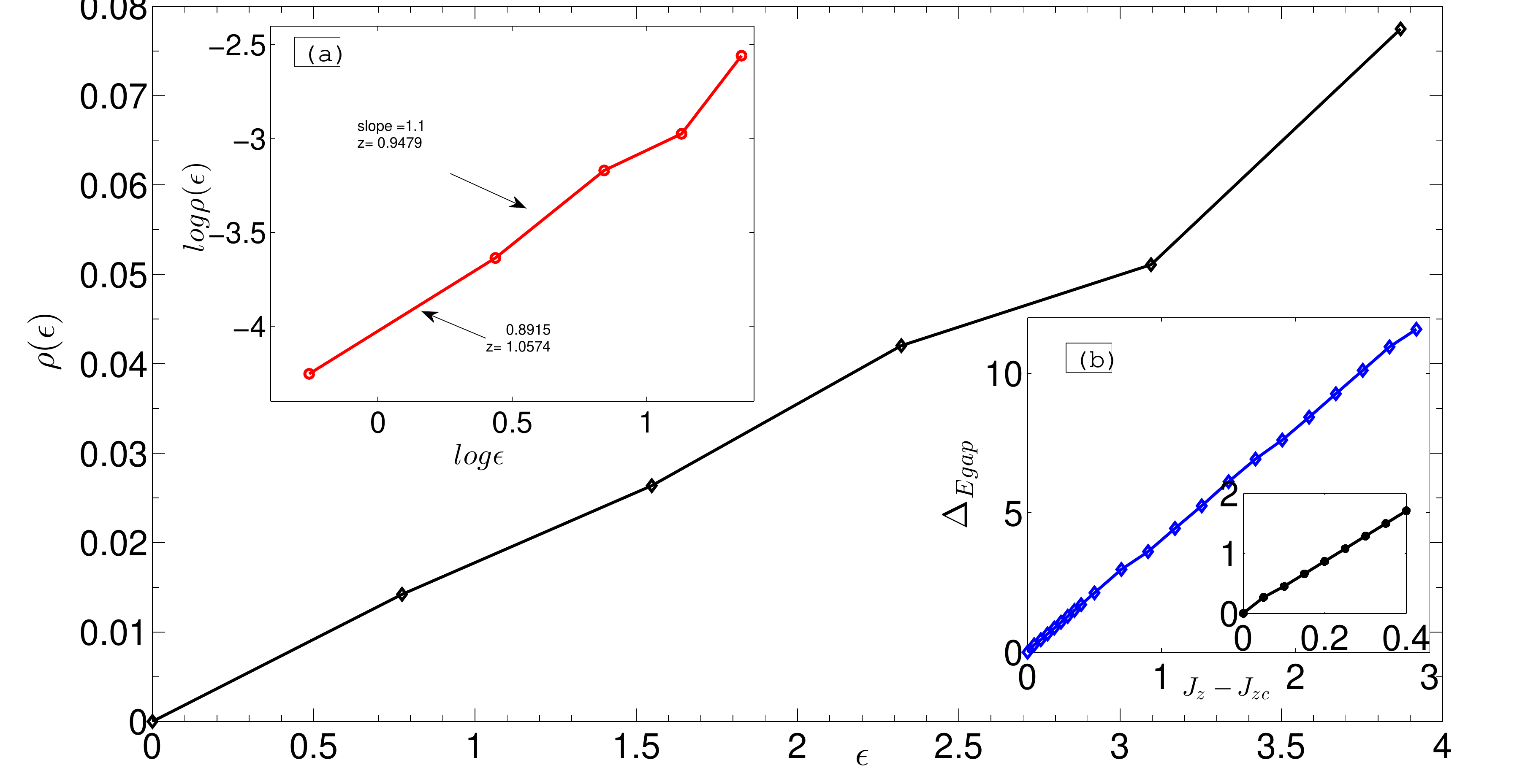,trim=0in 0in 0in 0.0in,clip=true,width=1.0\linewidth}}\hspace{-0.0\linewidth}
\caption{(Color online) Main panel showing the energy-dependence of the local density-of-states of the $c$-fermions at the ``Mott'' critical point. Using the scaling relation (see SI(2)), $\rho(\epsilon)\simeq \epsilon^{(d/z)-1)}$, we deduce $z=1$. The inset shows the extraction of the correlation length exponent $\nu=1$, using $E_{gap}^{Mott}(J_{z})\simeq (J_{z}-J_{z}^{(c)})^{z\nu}$ with $J_{z}^{(c)}= 2.14$. These estimates show that the gapless-QSL to gapped-QSL transition at finite $T$ falls into the $d=2$ Ising universality class.}
\label{fig3}
\end{figure}

   We now investigate the universality class of the continuous ``Mott'' transition found above. This is accomplished by estimating the critical exponents $\nu$ and $z$, related respectively to the critical divergences of the spatial and temporal correlation lengths at a continuous phase transition. In Fig.~\ref{fig3}, we show how the ``Mott'' energy gap, varies as a function of $J_{z}$ and then from $E_{g} \simeq (J_{z}-J_{z}^{c})^{z\nu}$, we estimate $z\nu=1$. Near a continuous Mott transition, the spatial correlations are controlled by a single length scale, the localization length $\xi$, which diverges at the MIT with a critical exponent $\nu$ as $\xi(J_{z})\simeq (J_{z}-J_{z}^{(c)})^{-\nu}$. To estimate $z$, we exploit the fact that the local density-of-states (LDOS) at the critical point must be independent of lattice size, and scales like $\rho(\omega)=|\omega|^{(d/z)-1}$ (see SI(2)). Using our numerical data (see Fig.~\ref{fig1}), we estimate $z=1.05$, taken equal to unity as the nearest approximation. Along with $z\nu=1$ and $z=1$, the correlation-length exponent at criticality is $\nu=1$. Remarkably, our values for $z$ and $\nu$ are precisely those characteristic of $d=2$ Ising criticality, putting the continuous Mott criticality we find in the $2d$ Ising universality class, in accord with early work~\cite{castellani} based on an approximate mapping.  Our finding of $z=1$ and $\nu=1$ implies that the magnetic Gr\"uneisen parameter will scale as $\Gamma_{m}(T)\simeq T^{-1/z\nu}=T^{-1}$~\cite{rosch-si} near the ``Mott'' QCP.

\begin{figure}
\centering
\includegraphics[angle=0,,trim=0.4in 1.4in 0in 1.0in,width=1.1\columnwidth]{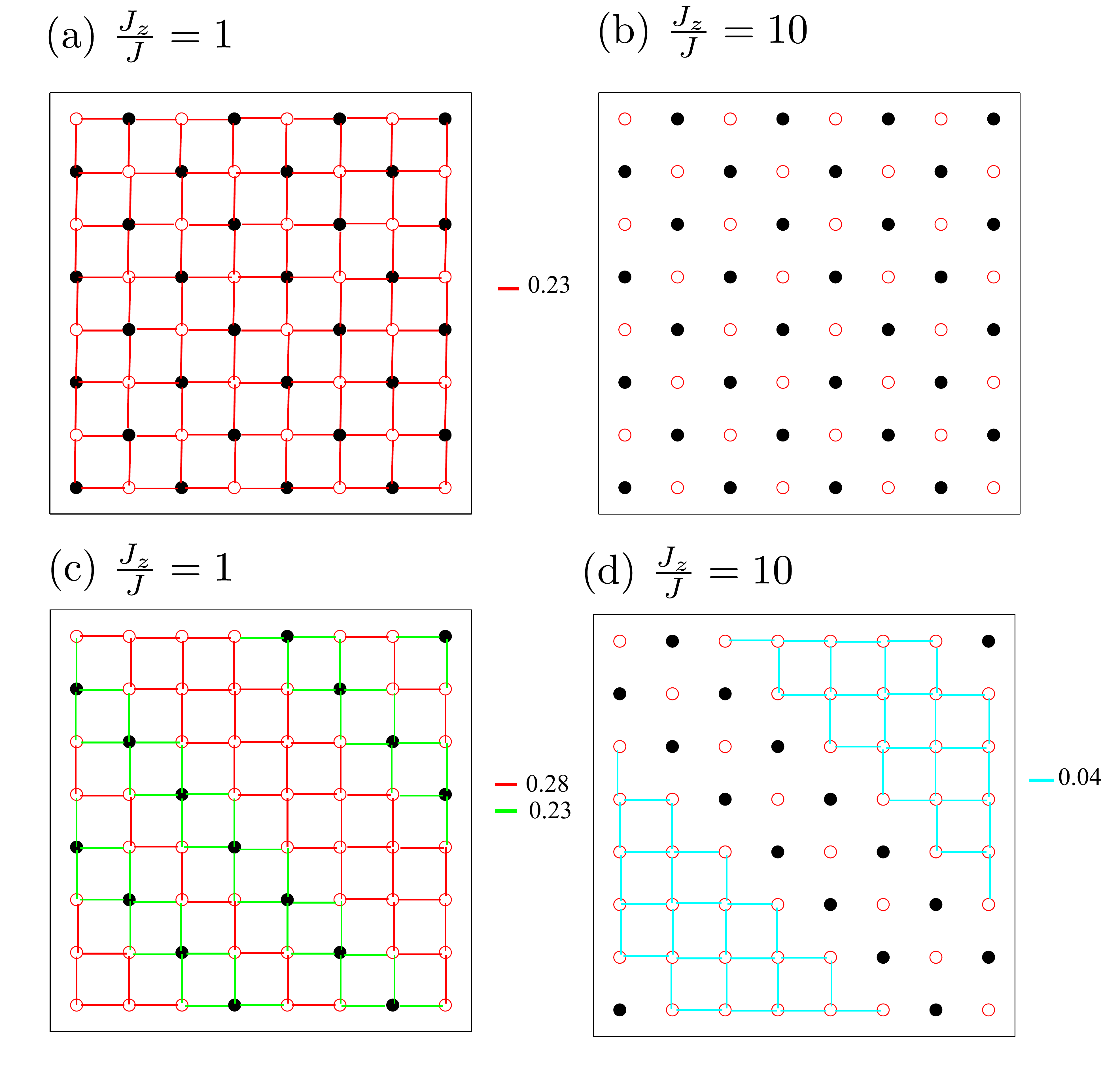}
\caption{(Color online) Flux ordered phases: Top panels (a) $J_z$=1, (b) $J_{z}$=10, show staggered flux order of the immobile $\alpha$-fermions for $n_{\alpha}/n_{c} = 1/2$, and for (c) $J_z$=1, (d) J$_z$=10, $n_{\alpha}/n_{c} = 1/4$. For $J_{z}=1$, the $p$-SC order is inhomogeneous for Case (c), being reduced on bonds linked to the fluxes (black circles). In Case (b), the Toric Code phase at $J_{z} = 10$ exhibits non-SC state, while in Case (d) diagonal stripe-order of fluxes, separated by regions with gossamer $p$-wave SC {\it without} flux obtains.}
\label{fig4}
\end{figure}

   Mapping to a spinless FKM also makes it clear that certain types of conventional order parameters are forbidden by Elitzur's theorem at finite $T$. Explicitly, local $Z_{2}$ gauge symmetry of $H_{FK}$ arising from $[n_{i\alpha},H_{FK}]=0$ for each ``site'' $i$ ( a $zz$ bond on the original honeycomb lattice) on the effective square lattice rules out ordered states corresponding to $\langle c_{i}^{\dag}\alpha_{i}\rangle, \langle c_{i}^{\dag}\alpha_{i}\alpha_{j}^{\dag}c_{j}\rangle$, etc, since these are not invariant under local $Z_{2}$ gauge symmetry. However, ``order'' corresponding to finite $\langle n_{ic}n_{jc}\rangle, \langle n_{i\alpha}n_{j\alpha}\rangle$ {\it are} invariant under local $Z_{2}$ gauge symmetry and, in fact, represent crystalline order in the usual FKM context~\cite{Kennedy}. Depending upon $n_{\alpha}/n_{c}$ at finite $T$ with $n_{c}+n_{\alpha}=1$, a wide range of ordered crystal phases are known to occur in that case. As representative examples, we present two such ``crystal'' states that occur in our case for $(i)$ $n_{c}=n_{\alpha}=1/2$ and $(ii)$ $n_{\alpha}/n_{c}=1/4$ in our FK model. In case $(i)$, we find that the $n_{\alpha}$ self-organize into a checkerboard ordered state, while a stripe ordered state obtains for case $(ii)$. That this does {\it not} imply any conventional symmetry breaking is bared by observing that an ordered crystal(s) of $\langle n_{\alpha}\rangle=\langle (1+ib_{i1}b_{i2})/2\rangle$ as above only means a staggered or stripe-ordered {\it flux} phase(s). In spin variables, such a flux condensate is a QSL without any conventional order~\cite{chen}. It is interesting that staggered ($\pi$)-flux phases were found to be competitively (compared to AF N\`eel state) stable in early mean-field studies of the $D=2$ Heisenberg model~\cite{affleck}. We find that $Z_{2}$-fluxes formed as bound states of dispersionless Majorana fermions ($ib_{i}b_{j}$ in Kitaev's fractionalization scheme, equivalently $n_{i\alpha}$ in $H_{FK}$) show staggered crystalline order in the featureless short-ranged~\cite{baskaran} spin liquid state.  This flux-ordering could explain the origin of the low-$T$ specific heat peak in the CDMFT~\cite{motome} study.  Interestingly, with {\it quenched} instead of annealed randomness, a disordered flux phase would obtain, where mesoscopic flux-rich regions would co-exist with flux-poor patches with local $p$-wave pairing: here, the JW fermions would propagate in a highly (intrinsically) disordered flux background in a way resembling the motion of fermions in a mesoscopically disordered system. At finite $T$, this would lead to an intrinsic glassy spin dynamics. Intentionally disordered $D=2$ honeycomb-based iridates do reveal smeared quantum criticality associated with such glassy~\cite{gegenwart} spin dynamics in a spin liquid: here, dopant impurities may act as an additional, quenched-disorder induced pinning centers, stabilizing such incipient glassy phase(s)~\cite{vojta}. 

   Our finding of a partially Gutzwiller projected BCS wavefunction above has deeper implications: it is very reminiscent of proposals in the context of QSL state(s) in $S=1/2$ Kagome antiferromagnets~\cite{hastings}, where strong evidence that the system is actually described by a Gutzwiller-projected wavefunction precisely of the form of our Eq.(3) is presented, with a Fermi sea being the Dirac Fermi sea for spinons with a $\pi$-flux through the hexagons. Precisely this is what happens for the finite-$T$ KM, where changing $n_{\alpha}$ from $+1$ to $-1$ on a single $zz$-link (a site ``$i$'' for the equivalent FK model on the effective sqaure lattice) induces a localized $\pi$-flux, but on a $zz$-bond sharing two neighbouring hexagons. Remarkably, this allows us to infer that the finite-$T$ QSL in the $d = 2$ KM is a {\it hidden Fermi liquid}~\cite{jain} of neutral spinons (or JW fermions). In our case, however, the JW fermions are coupled to an annealed $Z_{2}$ gauge field, which also enforces the constraint of a single JW fermion per site by the Gutzwiller projection for large $J_{z}$. Viewed from the perspective of ``hidden'' fermions, this hidden-FL of neutral JW-fermions has an appealing interpretation of the QSL and TC phases as partially and fully Gutzwiller-projected $p$-wave BCS states, but a much less transparent structure of highly non-local string condensates when viewed in terms of the original spins, much as in the famed cases of the fractional quantum Hall effect and high-$T_{c}$ cuprates~\cite{jain}. These emergent connections should aid in description of the spin liquid and ordered phases of perturbed Kitaev models using partial- or fully Gutzwiller-projected wavefunctions in numerical studies, an enterprise which has been very successful in studies of $d=2$ Heisenberg models~\cite{sorella}.  Finally, our approach can also be used to study the $d=3$ Kitaev models~\cite{nasu,trebst2} and realizations of Kitaev QSL in metal-organic frameworks: the former admit JW (or Majorana) Fermi surfaces and/or Weyl nodes, opening the possibility of having novel phase transitions between such states and ``Mott'' phase(s).  These aspects are deferred for future work.

\vspace{0.20cm}

{\bf Acknowledgement}

We thank M. Hermanns and G. Baskaran for discussions and the MPIPKS, Dresden for support whilst part of the work was done.

\clearpage

\pagebreak
\widetext
\twocolumngrid
\begin{center}
\textbf{\large Supplementary Informations}
\end{center}
\setcounter{equation}{0}
\setcounter{figure}{0}
\setcounter{table}{0}
\setcounter{page}{1}
\makeatletter
\renewcommand{\theequation}{S\arabic{equation}}
\renewcommand{\thefigure}{S\arabic{figure}}

{\bf (1) $d = 2$ Duality Transformation}

 Kitaev's famed solution relies on mapping spin-$1/2$s to bilinear of
four $(b^{x},b^{y},b^{z},c)$ Majorana fermions along the three ($xx,yy,zz$) bonds (Fig.~\ref{fig4}). We start with an alternative but equivalent approach due to Chen and Hu~\cite{chen} and Nussinov~\cite{nussinov} that exploits a duality mapping of the spins via a $d=2$ Jordan-Wigner (JW) fermionization of $H$. Remarkably, since this affords an {\it exact} mapping between topological and classical orders in $d = 2$ spin
systems, studying TO(s), their excitations and instabilities is transmuted into the better-known language of well-known models hosting classical order(s) characterized by symmetry-breaking. Thus,

\vspace{0.1em}

\be
S_{ij}^{+}=2[\Pi_{j'<j,i} S_{i'j'^{z}}] [\Pi_{i'<i} S_{i'j}^{z}] c_{ij}^{\dag}
\ee
and

\be
S_{ij}^{z}=(2c_{ij}^{\dag}c_{ij}-1)
\ee
and defining Majorana fermions on the two (``black'' and ``white'') sublattice sites of the honeycomb or brick-wall lattice as
$A_{w}=(c-c^{\dag})_{w}/i,B_{w}=(c+c^{\dag})_{w}$ and $A_{b}=(c+c^{\dag})_{b}, B_{b}=(c-c^{\dag})_{b}/i$, followed by the introduction of fermions $c=(A_{w}+iA_{b})/2,c^{\dag}=(A_{w}-iA_{b})/2$.  Using these transformations, we get the fermionized form of $H=H_{K}$ used in the main text.  In $H$ as thus defined in the main text, $\alpha_{i}'=iB_{i1}B_{i2}$ is a bilinear of Majorana fermions.  Since $[\alpha_{i}', H]=0$ for all $i$, it takes eigenvalues $\pm 1$. Thus, we define $\alpha_{i}'=(2n_{i,\alpha}-1)$, where $n_{i,\alpha}=0,1$ and $[n_{i,\alpha},H]=0$ for all $i$ as well, preserving the local Z$_{2}$ symmetry. At $T=0$, the flux-free condition, wherein {\it all} $\alpha_{i}'=+1$ (or $-1$), $H_{K}$ is exactly solvable.  With the additional FK-like term involving annealed $Z_{2}$ fluxes interacting with $c$-fermions, there is no exact solution in
$d=2,3$.

\vspace{0.20cm}

{\bf (2)  Scaling form of the DOS at the ``Mott'' Transition of JW Fermions}

   To unearth the scaling form of the local density-of-states (LDOS) that we use to derive $d=2$ Ising critical exponents at the ``Mott'' transition of the JW fermions in the main text, we start with the number of states, $N(\omega,L)$ below energy $\omega$ in a system of size $L$ in $d = 2$. This is a function of two dimensionless parameters, $\xi/L$ and $\omega/\Omega$,

\be
N(\omega,L)=G(L/\xi,\omega/\Omega)
\ee
Here, $\xi$ is the characteristic length scale and $\Omega$ a characteristic energy scale, related to each other via the dynamical critical exponent, $z$, as $\Omega\simeq \xi^{-z}$. Since $N(\omega,L)$ must scale as $L^{d}$,
this implies that

\be
N(\omega,L)=(L/\xi)^{d}g(\omega\xi^{z})
\ee
The LDOS is directly computed as 

\be
\rho(\omega)=L^{-d}\frac{dN}{d\omega}
\ee
leading to the scaling relation

\be
\rho(\omega)=\rho(-\omega)=\xi^{z-d}g'(|\omega|\xi^{z})
\ee
where we have invoked ``particle-hole'' symmetry about $\omega=0$ in the numerical LDOS (see Fig.~\ref{fig1} in the main text). Introducing a distance from the ``Mott'' QCP as $\delta=|J_{z}-J_{z}^{(c)}|/J_{z}^{(c)}$, and assuming that $\xi(J_{z})$ diverges like $\xi(J_{z})]\simeq \delta^{-\nu}$ near the QCP, with $\nu$ being the correlation length exponent, we find that 

\be
\rho(\omega)\simeq \delta^{(d-z)\nu}g'(|\omega|\delta^{-z\nu})
\ee
Finally, since one has a Dirac-like spectrum in the fermionized KM near the QCP, we expect that the LDOS will vary like $\rho(\omega)\simeq |\omega|^{d-1}$ in $d$ dimensions. Right at the QCP, since the $\xi$-dependence of the LDOS must vanish, we deduce that 

\be
\rho(\omega)\simeq \delta^{(d-z)\nu}(|\omega|\delta^{-z\nu})^{(d-z)/z}=|\omega|^{(d-z)/z}
\ee
This is the scaling form we have used close to the QCP to extract the dynamical critical exponent $z$. Using our MC+ED results, we thus find $z=1$, as stated in the main text.


\begin{thebibliography}{50}
\bibitem{kitaev} A. Kitaev, Annals of Physics 303, 2 (2003), ISSN 0003-4916.†
\bibitem{jilla} A. Buhler et al., Nature Communications 5 (2014).
\bibitem{jja} J. Q. You et al., Phys. Rev. B 81, 014505 (2010).
\bibitem{khaliullin} G. Jackeli and G. Khaliullin, Phys. Rev. Lett. 102, 017205 (2009).
\bibitem{trebst1} J. Reuther, R. Thomale, and S. Trebst, Phys. Rev. B 84, 100406 (2011).
\bibitem{moessner} J. Knolle et al., Phys. Rev. Lett. 112, 207203 (2014).
\bibitem{motome} J. Yoshitake, J. Nasu and Y.Motome, Phys. Rev. Lett. 117, 157203 (2016); ibid 113, 197205 (2014)
\bibitem{affleck} J. B. Marston and I. Affleck, Phys. Rev. B {\bf 39}, 11538 (1989).
\bibitem{Taraphder} U. K. Yadav {\it et al}., J. Phys: Condens Matter {\bf 22}, 5602 (2010).
\bibitem{Kennedy} T. Kennedy, Reviews in Mathematical Physics 06, 901 (1994).
\bibitem{nussinov} Han-Dong Chen and Zohar Nussinov, J. Phys. A. {\bf 41} 075001 (2008).
\bibitem{sorella} S. Yunoki and S. Sorella, Phys. Rev. Lett. 92, 157003 (2004).
\bibitem{wen} See ``Quantum Field Theory of Many-Body Systems'', X-G Wen, Oxford Graduate Texts (2007).
\bibitem{hu} H.-D. Chen and J. Hu, Phys. Rev. B 76, 193101 (2007).
\bibitem{fradkin} Z. Nussinov and E. Fradkin, Phys. Rev. B 71, 195120 (2005).
\bibitem{science1987} P. W. Anderson, Science {\bf 235}, 1196 (1987).
\bibitem{castellani} C. Castellani, et al., Phys. Rev. Lett. {\bf 43}, 1957 (1979).
\bibitem{rosch-si} Lijun Zhu {\it et al}., Phys. Rev. Lett. {\bf 91}, 066404 (2003).
\bibitem{baskaran} G. Baskaran, S. Mandal, and R. Shankar, Phys. Rev. Lett. 98, 247201 (2007).
is seen there for both one- and multi-orbital Hubbard models, especially with finite crystal-field terms.
\bibitem{gegenwart} S. Manni, Y. Tokiwa and P. Gegenwart, Phys Rev. B{\bf 89}, 241102(R) (2014).
\bibitem{hastings} M. B. Hastings, Phys Rev B {\bf 63}, 041143 (2000); ibid Y. Iqbal {\it et al.}, New J. Phys. {\bf 14}, 115031 (2012).
\bibitem{jain} J. Jain and P. W. Anderson, PNAS {\bf 106}, 9131 (2009).
\bibitem{vojta} M. Vojta, Phil. Mag., {\bf 86}, 1807 (2006)
\bibitem{nasu}  J. Nasu {\it et al.}, Phys Rev B {\bf 89}, 115125 (2014).
\bibitem{trebst2} S. Trebst, arXiv:1701.07056, and references therein.
\end{thebibliography}

\vspace{0.20cm}




\vspace{0.50cm}
\vspace{0.1em}


\vspace{0.20cm}



  
\begin{thebibliography}{20}
\bibitem{chen} H.-D. Chen and J. Hu, Phys. Rev. B 76, 193101 (2007).
\bibitem{nussinov} H-D. Chen and Z. Nussinov, Journal of Physics A: Mathematical and Theoretical {\bf 41} (2008).

\end{thebibliography}
\end{document}